\documentclass[manuscript]{aastex}

\begin{document}
\pagenumbering{arabic}

\title{ LUMINOSITIES OF BARRED AND UNBARRED S0 GALAXIES}

\author{Sidney van den Bergh}
\affil{Dominion Astrophysical Observatory, Herzberg Institute of Astrophysics, National Research Council of Canada, 5071 West Saanich Road, Victoria, BC, V9E 2E7, Canada}
\email{sidney.vandenbergh@nrc.gc.ca}

\begin{abstract}

 Lenticular galaxies with  $M_{B} < -21.5$ are almost exclusively
unbarred, whereas both barred and unbarred objects occur at fainter luminosity 
levels. This effect is observed both for objects classified in blue light, and 
for those that were classified in the infrared. This result suggests that the 
most luminous (massive) S0 galaxies find it difficult to form bars. As a 
result the mean luminosity of unbarred lenticular galaxies in both B and IR 
light is observed to be $\sim$0.4 mag brighter than than that of barred 
lenticulars. A small contribution to the observed luminosity difference that 
is found between SA0 and SB0 galaxies may also be due to the fact that there 
is an asymmetry between the effects of small classification errors on SA0 and 
SB0 galaxies. An E galaxy might be misclassified as an S0, or an S0 as an E. 
However, an E will never be misclassified an SB0, nor will an SB0 ever be 
called an E. This asymmetry is important because elliptical (E) galaxies are 
typically twice as luminous as lenticular (S0) galaxies. The present results 
suggest that the evolution of luminous lenticular galaxies may be closely 
linked to that of elliptical galaxies, whereas fainter lenticulars might be 
more closely associated with ram-pressure stripped spiral galaxies. Finally it 
is pointed out that fine details of the galaxy formation process might account 
for some of the differences between the classifications of the same galaxy by 
individual competent morphologists.
\end{abstract}

\keywords{galaxies: elliptical and lenticular}

\section{INTRODUCTION}

 The class of lenticular (S0) galaxies was introduced by Hubble
(1936) as a more-or-less speculative means of bridging the morphological chasm 
between elliptical (E) and spiral (S) galaxies. The definition of the S0 class 
was later improved and expanded by de Vaucouleurs (1959) and by Sandage 
(1961). Since then numerous individual lenticular galaxies have been 
classified on large-scale photographic plates by Sandage \&  Tammann (1981), 
Sandage \& Bedke (1994), de Vaucouleurs et al. 
(1991) and by Buta, Corwin \& Odewahn (2007). Inter-comparison of the 
individual classifications of lenticular galaxies by these expert 
morphologists reveals so much dispersion that King (1992) and Djorgovski 
(1992) proposed giving up entirely on optical morphological classification and 
replacing it by a system of measured physical parameters. A less drastic, and 
perhaps more productive, procedure has been proposed by Laurikainen et 
al. (2011, 2012), who obtained large diameter images at a wavelength of 2.2 
microns which trace the old stellar populations of early-type galaxies in a 
fashion that is essentially free of the influence of internal extinction and 
the effects of recent star formation. It is the purpose of the present paper 
to use these infrared morphological classifications of S0 galaxies by 
Laurikainen et al., in conjunction with their distances and luminosities 
assigned by Sandage \& Tammann (1981) [who assumed $H_{o}$ = 50 km sec$^{-1}$  Mpc$^{-1}]$ to study the properties of the bright relatively 
nearby lenticular galaxies. For the sake of convenience the $M^{o,i} _{B_{T}}$ 
magnitudes of Sandage \&  Tammann (1981) will subsequently be referred to as $M_T$ 
The luminosity {\it differences} between various classes of galaxies that are 
discussed in the present paper are, of course, independent of the value of the 
Hubble parameter adopted by Sandage\&  Tammann.

\section{LUMINOSITIES OF UNBARRED AND BARRED S0 GALAXIES}

 Laurikainen et al. (2011) have recently published a near-IR
atlas of early-type galaxies. Their data are discussed in more detail by
Laurikainen et al. (2012). Table 3 of the former paper lists galaxy 
classifications made on the basis of inspection of 2.2 micron galaxy images. 
The luminosity distributions of non-barred and barred early-type galaxies in 
their data are listed in Table 1. Inspection of this table shows that non-barred (SA0) galaxies are typically more luminous than barred (SB0) galaxies. 
Galaxies which these authors assign to the intermediate class SAB0 are found 
to be of intermediate luminosity. A Kolmogorov-Smirnov test shows that there 
is only a 2\% probability that the luminosity distributions of SA0 and of SB0 
galaxies in this table were drawn from the same parent population. Table 2 
shows that the difference between the frequency distribution of barred and 
unbarred lenticular galaxies is almost entirely due to the absence of
SB0 galaxies that are more luminous than $M_B = -21.5$. Table 3 shows a similar 
effect for the galaxies classified in blue light by Sandage \& Tammann. [It is 
noted in passing that both of the two most luminous SB0 galaxies in the 
Revised Shapley-Ames Catalog have $M_{B} = -21.51$.]  Inter comparison of Tables 2 
and 3 shows that the almost complete absence of luminous barred lenticular 
galaxies is observed in both blue and infrared light. It is also noted that 
the difference between the median magnitudes of barred and unbarred lenticular 
galaxies is $\sim$0.4 mag in both blue and infrared light. In other words the 
absence (or near absence) of luminous barred S0 galaxies is a feature that 
occurs in both blue light and in infrared classifications.  A possible 
contribution to the small observed difference between the median luminosities 
of SA0 and
SB0 galaxies might be provided by the asymmetry in the effects of small 
classification errors that results from the fact that there are no barred 
ellipticals. As a result an E galaxy might be slightly misclassified as an SA0 
and an SA0 might be misclassified as an E. 
However, because there are no barred ellipticals, an SB0 will never be 
classified as an E.

It is important to always keep in mind that the classification of
early-type galaxies represents a considerable challenge to precision 
morphology. This is shown most clearly by the rather large scatter between the 
classification types assigned to the same galaxies by different competent 
morphologists. Table 4 shows a comparison between the Laurikainen and Sandage 
\& Tammann classifications of early-type galaxies with Sandage luminosities 
greater than $M_{B}$ = -21.5. The relatively low degree of agreement between these 
classification types is disappointing. However, for objects fainter than  $M_{B}$ = 
-21.5 the agreement appears better with $\sim$2/3 of all galaxies being assigned to 
the same morphological types by these two sets of authors. It seems that we 
may have approached the outer boundary of the applicability of morphological 
classification, where ``noise'' between individual classifications (and 
classification systems) becomes a non-negligible factor. Such noise may arise 
from factors such as differences in display technique (inspection of plates 
versus computer monitors and NICMOS arrays) or from the presence of subtle inner lenses 
or shallow core profiles. No two galaxies are identical, so one should not 
expect all such objects to exhibit the features that define their 
morphological types with the same strength or in exactly the same way. In this 
connection it is of interest to note that Sales et al. (2012) have used 
detailed modeling to show that most spheroidal galaxies consist of 
superpositions of stellar components with distinct kinematics, ages and 
metallicities, an arrangement that might survive to the present day because of 
the paucity of recent major mergers. In particular cold inflows of gas along 
separate filaments with misaligned spins might settle on off-axis orbits 
relative to material that had been accreted earlier. The detailed history of 
gas inflow might therefore affect disk and core formation in ways that lead to 
small systematic differences between the ways in which these objects are 
classified by different expert morphologists.

\section{LUMINOSITIES OF ELLIPTICAL GALAXIES}

Only a small number of galaxies that were classified in the IR
by Laurikainen et al. are ellipticals. These objects are typically found to be 
more luminous that those assigned to type S0. This agrees with previous 
results by van den Bergh (1998, p.61) and van den Bergh
(2011) which clearly show that S0 galaxies are systematically less luminous 
than either E or Sa galaxies - thus contradicting Hubble's notion that S0 
galaxies are truly intermediate between Hubble stages E and Sa. Ellipticals 
are typically $\sim$2 times more luminous than lenticulars. As a result E galaxies 
misclassified as S0s will increase the mean luminosity of the SA0 sample. 
However, the mean luminosity of
SB0 will be unaffected because elliptical galaxies do not have bars and 
therefore cannot be misclassified as being of type SB0.

\section{DISCUSSION}

The principal results obtained in the present investigation is that
luminous unbarred lenticular galaxies are common, whereas such luminous 
objects  are rare or absent among barred lenticulars. This conclusion is found 
to hold for both galaxies classified in the blue (Sandage \& Tammann 1981) and 
for those classified in the infrared by Laurikainen et al. (2011). Since 
elliptical galaxies are, on average, known to be twice as luminous as 
lenticulars this suggests a possible evolutionary connection between 
elliptical galaxies and luminous lenticulars. On the other hand the lower 
luminosity barred lenticular galaxies may, from an evolutionary point of view, 
be more closely related to spiral galaxies that have been stripped of gas by 
ram-pressure. A small contribution to the observed $\sim$0.4 mag. mean luminosity 
difference between barred and unbarred lenticular galaxies might also be due 
to the fact that there is an asymmetry in the effects of small morphological 
classification errors, which results from the fact that there are no barred 
elliptical galaxies. As a result an E galaxy might be misclassified as an SA0, 
but an SB0 would never be misclassified as an E.  The effects of gas accretion on bar formation (Bournaud \& Combes 2002) are probably not significant for very early-type galaxies, such as those of type SBO.

 It is a pleasure to acknowledge helpful exchanges of e-mail with
Bob Abraham, Eija Laurikainen, Heikki Salo, Ron Buta and Johan Knapen.  I also thank a particularly kind referee.

\begin{deluxetable}{lccc}
\small
\tablewidth{0pt}
\tablecaption{ Luminosity distribution of lenticular galaxies in the catalog of Laurikainen et al.} 

\tablehead{\colhead{$M_{B}$} & \colhead{N(SA0)} &  \colhead{N(SAB0)} &  \colhead{N(SB0)}}

\startdata

-22.50 to -22.99   &   1  &      0    &    0\\
-22.00 to -22.49   &   5   &    1      &  0\\
-21.50 to -21.99    &    9  &     2    &    0\\
-21.00 to -21.49   &   6  &      4  &      5\\
-20.50 to -20.99   &    6   &    6   &      5\\
-20.00 to -20.49   &   3   &    7   &     7\\
-19.50 to -19.99   &   7    &   3    &     8\\
-19.00 to -19.49   &   3   &    1    &    4\\
-18.50 to -18.99   &   0   &    0    &    2\\
$ >$  -18.50     &       1   &   0   &    0\\

\enddata
\normalsize
\end{deluxetable}

\begin{deluxetable}{lcll}
\small
\tablewidth{0pt}
\tablecaption{  Luminosity distribution of SA0, SAB0 and SB0 galaxies  (Laurikainen et al.2011) classified in the infrared.} 

\tablehead{\colhead{  }         & \colhead{N(SA0)}             &  \colhead{N(SAB0)}         &  \colhead{N(SB0)}}

\startdata

Bright ($M_{B} < $ -21.5)   &   15    &    1      &   0\\
Faint  ($M_{B} > $ -21.5)    &  26    &    23    &   31\\

\enddata
\normalsize
\end{deluxetable}

\begin{deluxetable}{ccc}
\small
\tablewidth{0pt}
\tablecaption{ Luminosity distribution of S0 and SB0 galaxies  classified in blue light (Sandage \&  Tammann 1981)} 

\tablehead{\colhead{  }         & \colhead{N(S0)}                 &  \colhead{N(SB0)}}

\startdata

Bright ($M_{B} < $  -21.5)    &    16    &     2\\
Faint  ($M_{B} > $  -21.5)     &     96   &    31\\

\enddata
\normalsize
\end{deluxetable}

\begin{deluxetable}{cccc}
\small
\tablewidth{0pt}
\tablecaption{  Comparison between infrared (Laurikainen et al) and blue (Sandage \&  Tammannn) classifications of luminous early-type galaxies with $M_{B} < $ -21.5} 

\tablehead{\colhead{S\&T type}      & \colhead{E}    & \colhead{E/S0}        &  \colhead{S0}}

\startdata

Laurikainen et al. classification &   &   & \\
SA0          &              4     &      1     &    7\\
SAB0       &               0    &      0      &  2\\
SB0           &            0      &  0          & 0\\

\enddata
\normalsize
\end{deluxetable}


\begin{references} 

\reference{}Barway, S., Wadadekar, Y. \&  Kembhavi, A. K. 2011,  MNRAS, 410, L18 
\reference{} Bournaud, F. \& Combes, F. 2002, A\&A, 392, 83
\reference{}Buta, R., Corwin, H. G., \&  Odewahn, S. C. 2007, The  de Vaucouleurs Atlas of Galaxies, Cambridge: Cambridge University Press.
\reference{} de Vaucouleurs, G. 1959, Handb. der Physik, 53, 275
\reference{}de Vaucouleurs, G., de Vaucouleurs, A., Corwin, H. G.,  Buta, R., Paturel, G. \&  Fouqu{\'e}, P. 1991,  Third Reference Catalogue of Bright Galaxies, New York: Springer 
\reference{}Djorgovski, S. 1992, in ``Morphology and Physical Classification of Galaxies'', eds. Longo, G., Capaccioli, M. \&  Busarello, G. Dordrecht: Kluwer Acad. p. 427
\reference{}Hubble, E. 1936, The Realm of the Nebulae, New Haven, Yale University Press, p. 45
\reference{}King, I. 1992, in ``Morphological and Physical Classification of Galaxies" eds. Longo, G., Cappaccioi, M. \&  Busarello,
 Dordrecht, Kluwer Acad. p. 371
\reference{}Laurikainen, E., Salo, H., Buta, R. \&  Knapen, J. H. 2011,  MNRAS, 418, 1452 
\reference{}Laurikainen, E., Salo, H., Buta, R. \&  Knapen, J. H. 2012,  Advances in Astronomy (in press = arXiv:1111.6447) 
\reference{}Sales, L. V., Navarro, J. F., Theuns, T., Schaye, J., White, S. D. M., Frenk, C. S., Crain, R. A. \&  Dalla Vecchia, C. 2012, MNRAS (in press = arXix:1112.2220) 
\reference{}Sandage, A. 1961, The Hubble Atlas of Galaxies, Washington:  Carnegie Institution of Washington Publ. No. 618 
\reference{} Sandage, A. \&  Bedke, J. 1994,  The Carnegie Atlas of Galaxies, Carnegie Institution of Washington Publ. No. 638
\reference{}Sandage, A. \&  Tammann, G. A. 1981, Revised Shapley-Ames Catalog of Bright Galaxies, Washington: Carnegie Institution of Washington
\reference{} van den Bergh, S. 1998, Galaxy Morphology and Classification,  Cambridg: Cambridge  University Press 
\reference{}van den Bergh, S. 2002, AJ, 124, 782 van den Bergh, S. 2011, PASP, 141,188

\end{references}
\end{document}